\documentclass[a4paper,11pt,tableheading]{scrartcl}
\pdfoutput=1

\usepackage[utf8]{inputenc}
\usepackage{authblk}
\usepackage{booktabs}
\usepackage{hyperref}
\usepackage{graphicx}

\usepackage{newtxtext}
\usepackage{newtxmath}

\usepackage{amsmath}
\usepackage{amssymb}
\usepackage{amsfonts}

\usepackage[font=small,labelfont=bf,labelsep=colon]{caption}
\usepackage[labelfont=up]{subcaption}

\usepackage[frozencache]{minted}
\setminted{frame=leftline,framesep=2mm,autogobble,fontsize=\small}

\usepackage[backend=biber,style=numeric-comp,sorting=none,giveninits=true]{biblatex}
\DeclareFieldFormat[article]{title}{\textit{#1}\isdot}
\DeclareFieldFormat[report]{title}{\textit{#1}\isdot}
\DeclareFieldFormat[booklet]{title}{\textit{#1}\isdot}
\DeclareFieldFormat[misc]{title}{\textit{#1}\isdot}

\DeclareFieldFormat[article]{journaltitle}{#1\isdot}
\DeclareFieldFormat[article]{journalsubtitle}{#1\isdot}
\DeclareFieldFormat[article]{volume}{\textbf{#1}\isdot}

\DeclareFieldFormat{pages}{\mkfirstpage[{\mkpageprefix[bookpagination]}]{#1}}

\renewcommand*{\newunitpunct}{\addcomma\space}
\renewbibmacro{in:}{}

\DefineBibliographyStrings{english}{%
    page={},
    pages={}
}

\usepackage{authblk}

\title{Extending the ARC Information Providers to report
information on GPU resources}

\author[1,a]{Max Isacson}
\author[1]{Mattias Ellert}
\author[1]{Richard Brenner}
\affil[1]{Uppsala University}
\affil[a]{\href{mailto:max.isacson@physics.uu.se}{max.isacson@physics.uu.se}}

\begin{document}

\maketitle

\section{Introduction}
General-purpose Computing on Graphics Processing Units (GPGPU) has been
introduced to many areas of scientific research such as
bioinformatics~\cite{schatz2007high}, cryptography~\cite{manavski2007cuda},
computer vision~\cite{pulli2012real}, and deep learning~\cite{Silver2016}.
However, computing models in the High-energy Physics (HEP) community are
still mainly centered around traditional CPU resources.  Tasks such as track fitting,
particle reconstruction, and Monte Carlo simulation could benefit greatly from
a high-throughput GPGPU computing model, streamlining bottlenecks in analysis
turnover. This technical note describes the basis of an implementation of an
integrated GPU discovery mechanism in GRID middleware to facilitate GPGPU.

The GRID~\cite{Bos:840543,Bird:1695401} computing model is defined by the
Worldwide LHC\footnote{Large Hadron Collider} Computing Grid (WLCG)
collaboration and comprises a worldwide network of geographically separated
GRID sites hosted by local organizations such as universities or computing and
data centres. The GRID is heterogeneous in nature since the resources provided
by the local organizations can be any combination of hardware and back-end
software. The Advanced Resource Connector~\cite{Ellert:2007faf} (ARC) middleware
was developed by NorduGrid~\cite{nordugrid} to combine several distributed resources
into a single entry point. Resources connected through ARC are presented
as a single virtual GRID site to the user. The ARC middleware takes the
responsibility of distributing a user submitted job specification to
the available connected resources.

Information about the local resources, or Compute Elements (CE), are gathered
by the ARC Information Providers with a central entry point provided by the
\texttt{CEinfo.pl} perl-script. The \texttt{CEinfo.pl} script determines the
Local Resource Management System (LRMS) flavour and calls the appropriate
perl-module which in turn queries the LRMS back-end. The collected information
is provided to the user either through the \texttt{arcinfo} command or the
Web-based GRID Monitor.

The modifications described in this note are done using the \texttt{ARC6RC1}
installed on the Kebnekaise cluster in Ume\aa\ provided by the
HPC2N~\cite{hpc2n} collaboration.  Kebnekaise uses the
\texttt{SLURM}~\cite{slurm} LRMS back-end and provides GPU resources through 80
NVIDIA Tesla K80 and and 20 NVIDIA Tesla V100 cards. The implementation is
described in Section~\ref{sec:impl}. A summary of all the modifications is
provided in Appendix~\ref{app:listings} and a complete set of \texttt{diff}s is
provided in Appendix~\ref{app:diffs}.

\section{Implementation}\label{sec:impl}
The \texttt{SLURM} back-end provides the \texttt{sinfo} command used to query the underlying
system. Nodes with GPU cores are listed as a
\emph{General Resource} which can be accessed using the
\texttt{-o "\%G"} flag. The output of \texttt{sinfo} on Kebnekaise is
\begin{minted}{bash}
    $ sinfo -aho "%G"
    (null)
    gpu:k80ce:4,mps:no_consume:1,gpuexcl:no_consume:1
    gpu:k80ce:8,mps:no_consume:1,gpuexcl:no_consume:1
    gpu:v100:2,mps:no_consume:1,gpuexcl:no_consume:1
    hbm:16G
    hbm:0
\end{minted}
where the \texttt{-a} flag is used to list all partitions and \texttt{-h} to
suppress the header.  The goal is to build a pipeline to provide this
information to the \texttt{arcinfo} command.  \texttt{SLURM} specific
information providers are implemented in the \texttt{SLURMmod.pm} module which
is extended to call and parse the appropriate \texttt{sinfo} command through
the \texttt{slurm\_read\_gresinfo()} routine. The array
\texttt{@sinfo\_gresinfo} is added to the \texttt{\$lrms\_cluster} table as
shown in Listing~\ref{lst:SLURMmod} and \ref{lst:LRMSInfo}.
\begin{minted}{perl}
sub slurm_read_gresinfo($){
  my @sinfo_gresinfo;
  my $gresinfo;
  open (SCPIPE,"$path/sinfo -a -h -o \"gresinfo=%G\"|");
  while(<SCPIPE>){
    my $string = $_;
    if ($string !~ m/\(null\)/) {
      $gresinfo = get_variable("gresinfo",$string);
      push(@sinfo_gresinfo, $gresinfo);
    }
  }
  close(SCPIPE);

  return @sinfo_gresinfo;
}
\end{minted}

With the information lifted out of the \texttt{SLURM} specific module it can be
accessed by other parts of the information system. The module
\texttt{ARC1ClusterInfo.pm} is responsible for collecting information from the
\texttt{LRMS} and converting it in into a format readable by the \texttt{XML}
printer. Inside the \texttt{ARC1ClusterInfo::collect} subroutine the gathered
General Resource information is appended to the Computing Manager table
\texttt{\$cmgr}
\begin{minted}{perl}
  $lrms_cluster->{gres} = [@sinfo_gresinfo];
\end{minted}
with details in Listing \ref{lst:ARC1ClusterInfo}. This table is
read by the \texttt{GLUE2xmlPrinter.pm} module and the output \texttt{XML}
document is extended with a \texttt{GeneralResources} node by
modifying the \texttt{ComputingManager()} and adding the
\texttt{GeneralResources()} subroutine, Listing~\ref{lst:GLUE2xmlPrinter}.
The \texttt{XML} output of this reads
\begin{minted}{xml}
<GeneralResources>
  <Resource>gpu:k80ce:4,mps:no_consume:1,gpuexcl:no_consume:1</Resource>
  <Resource>gpu:k80ce:8,mps:no_consume:1,gpuexcl:no_consume:1</Resource>
  <Resource>gpu:v100:2,mps:no_consume:1,gpuexcl:no_consume:1</Resource>
  <Resource>hbm:16G</Resource>
  <Resource>hbm:0</Resource>
</GeneralResources>
\end{minted}
which can be parsed. A sample of the \texttt{XML} output with parent nodes
included is shown in Listing~\ref{lst:XMLoutput}.

Endpoints can be queried using the \texttt{arcinfo} command which reads the
\texttt{XML} output and formats it into a human readable format. To make
\texttt{arcinfo} aware of the General Resources information a new data
member
\begin{minted}{C++}
std::list<std::string> GeneralResources;
\end{minted}
is added to the
\texttt{Arc::ComputingManagerAttributes} class as shown in Listing
\ref{lst:ExecutionTarget.h} and the stream
operator is modified, as shown in Listing \ref{lst:ExecutionTarget.cpp}, to
make the data printable. The data members
are filled from the \texttt{XML} output inside the
\texttt{GLUE2::ParseExecutionTargets}
subroutine which is modified to parse the supplemented information in the
\texttt{XML} tree as shown in Listing
\ref{lst:GLUE2.cpp}. The output from the \texttt{arcinfo} command now reads
\begin{minted}{bash}
Computing service:
  # ...
  Batch System Information:
    # ...
    General resources:
      gpu:k80ce:4,mps:no_consume:1,gpuexcl:no_consume:1
      gpu:k80ce:8,mps:no_consume:1,gpuexcl:no_consume:1
      gpu:v100:2,mps:no_consume:1,gpuexcl:no_consume:1
      hbm:16G
      hbm:0

    # ...
  # ...
\end{minted}
showing that the requested information is properly propagated through the
pipeline.

To utilize the GPU cores a specialized Runtime Environment (RTE)
has to be supplied to configure the job submission script.
A minimal example is shown in Listing
\ref{lst:rte} where the environment variable
\texttt{joboption\_nodeproperty\_\#} is modified to append
the option \texttt{\#SBATCH --gres=gpu:k80:1} to the \texttt{SLURM}
job specification so that the requested GPU cores are allocated, in
this example a single K80 card. A
minimal job description in the \texttt{XRSL}~\cite{xrsl}
language is shown in Listing \ref{lst:xrsl}
where the new RTE is requested to access the
GPU cores. The \texttt{XRSL} file is submitted with the
\texttt{arcsub} command.

\begin{listing}[H]
\begin{minted}{bash}
#!/bin/bash

case "$1" in
    0) # called during creation of batch script on frontend
        export joboption_rsl_project=SNIC20XX-Y-ZZ
        export joboption_nodeproperty_0="--gres=gpu:k80:1"
        ;;
    1) # called before execution of the main executable on the computing node
        module load GCC/6.4.0-2.28
        module load PGI/17.10-GCC-6.4.0-2.28
        module load CUDA/9.0.176
        ;;
    2) # called after execution of the main executable on the computing node
        ;;
    *) # error
        return 1
        ;;
esac
\end{minted}
\caption{Minimal runtime environment for a GPU job.}\label{lst:rte}
\end{listing}

\begin{listing}[H]
\begin{minted}{text}
&
(jobName="MinimalGpuJob")
(executable="some_executable")
(runTimeEnvironment="ENV/KGPU")
(inputFiles=("some_executable" "/path/to/some_executable"))
(outputFiles=("/" ""))
(wallTime="30")
(stdout="std.out")
(stderr="std.err")
\end{minted}
\caption{Minimal \texttt{XRSL} job specification for a GPU job.}\label{lst:xrsl}
\end{listing}

\section*{Acknowledgements}
The authors would like to thank the Swedish National Infrastructure for
Computing (project SNIC2019-5-39) for providing computing resources.

\newpage
\printbibliography[heading=bibintoc]

\newpage
\appendix
\section{Listings summarising all modifications}\label{app:listings}

\begin{listing}[H]
\begin{minted}{perl}
our(/*...*/, @sinfo_gresinfo);

#...

sub cluster_info() {
  # ...
  $lrms_cluster->{gres} = [@sinfo_gresinfo];
  # ...
}

sub slurm_get_data {) {
  # ...
  @sinfo_gresinfo = slurm_read_gresinfo();
}

# ...

sub slurm_read_gresinfo($){
  my @sinfo_gresinfo;
  my $gresinfo;
  open (SCPIPE,"$path/sinfo -a -h -o \"gresinfo=%G\"|");
  while(<SCPIPE>){
    my $string = $_;
    if ($string !~ m/\(null\)/) {
      $gresinfo = get_variable("gresinfo",$string);
      push(@sinfo_gresinfo, $gresinfo);
    }
  }
  close(SCPIPE);

  return @sinfo_gresinfo;
}
\end{minted}
\caption{Modifications and additions in the \texttt{SLURMmod.pm} module.}\label{lst:SLURMmod}
\end{listing}

\begin{listing}[H]
\begin{minted}{perl}
my $lrms_info_schema = {
  'cluster' => {
  # ...
    'gres' => ''
  },
  # ...
};
\end{minted}
\caption{Modifications and additions in the \texttt{LRMSInfo.pm} module.}\label{lst:LRMSInfo}
\end{listing}

\begin{listing}[H]
\begin{minted}{perl}
sub collect($) {
  # ...
  my $getComputingService = sub {
    # ...
    my $getComputingManager = sub {
      # ...
      $cmgr->{GeneralResources}{Resource} = $cluster_info->{gres};
      # ...
    };
    # ...
  };
  # ...
}
\end{minted}
\caption{Modifications and additions in the \texttt{ARC1ClusterInfo.pm} module.}\label{lst:ARC1ClusterInfo}
\end{listing}

\begin{listing}[H]
\begin{minted}{perl}
sub ComputingManager {
  Element(@_, 'ComputingManager', 'Manager', sub {
    # ...
    $self->begin('GeneralResources');
    $self->GeneralResources($data->{GeneralResources});
    $self->end('GeneralResources');
    # ...
  });
}

# ...

sub GeneralResources {
  my ($self, $data) = @_;
  $self->properties($data, 'Resource');
}
\end{minted}
\caption{Modifications and additions in the \texttt{GLUE2xmlPrinter.pm} module.}\label{lst:GLUE2xmlPrinter}
\end{listing}

\begin{listing}[H]
\begin{minted}{C++}
namespace Arc {
/* ... */
  class ComputingManagerAttributes {
  public:
    /* ... */
    std::list<std::string> GeneralResources;
    /* ... */
  };
  /* ... */
}
\end{minted}
\caption{Modifications and additions in the \texttt{ComputingManagerAttributes}
class in \texttt{ExecutionTarget.h}.}\label{lst:ExecutionTarget.h}
\end{listing}

\begin{listing}[H]
\begin{minted}{C++}
namespace Arc {
  /* ... */
  std::ostream operator<<(std::ostream& out,
      const ComputingManagerAttributes cm) {
    /* ... */
    if (!cm.GeneralResources.empty()) {
      out << IString("General resources:") << std::endl;
      for (std::list<std::string>::const_iterator it =
          cm.GeneralResources.begin();
          it != cm.GeneralResources.end(); ++it) {
        out << "  " << *it << std::endl;
      }
    }
    /* ... */
  }
  /* ... */
}
\end{minted}
\caption{Modifications to the stream operator for the \texttt{ComputingManagerAttributes}
class in \texttt{ExecutionTarget.cpp}.}\label{lst:ExecutionTarget.cpp}
\end{listing}

\begin{listing}[H]
\begin{minted}{C++}
namespace Arc {
  /* ... */
  void GLUE2::ParseExecutionTargets(XMLNode glue2tree,
      std::list<ComputingServiceType>& targets) {
    /* ... */
    for (; GLUEService; ++GLUEService) {
      /* ... */
      for (XMLNode xComputingManager = GLUEService["ComputingManager"];
          (bool)xComputingManager; ++xComputingManager) {
        /* ... */
        if (xComputingManager["GeneralResources"]) {
          for (XMLNode n = xComputingManager["GeneralResources"]["Resource"];
              n; ++n) {
            ComputingManager->GeneralResources.push_back((std::string)n);
          }
        }
        /* ... */
      }
      /* ... */
    }
    /* ... */
  }
  /* ... */
}
\end{minted}
\caption{Modifications of the \texttt{GLUE2} class in \texttt{GLUE2.cpp}.}
\label{lst:GLUE2.cpp}
\end{listing}

\begin{listing}
\begin{minted}{xml}
<InfoRoot>
  <Domains>
    <AdminDomain>
      <!-- ... -->
      <Services>
        <ComputingService>
          <!-- ... -->
          <ComputingManager>
            <!-- ... -->
            <GeneralResources>
              <Resource>gpu:k80ce:4,mps:no_consume:1,gpuexcl:no_consume:1</Resource>
              <Resource>gpu:k80ce:8,mps:no_consume:1,gpuexcl:no_consume:1</Resource>
              <Resource>gpu:v100:2,mps:no_consume:1,gpuexcl:no_consume:1</Resource>
              <Resource>hbm:16G</Resource>
              <Resource>hbm:0</Resource>
            </GeneralResources>
            <!-- ... -->
          </ComputingManager>
        <!-- ... -->
        </ComputingService>
      </Services>
    </AdminDomain>
  </Domains>
</InfoRoot>
\end{minted}
\caption{Output sample of the \texttt{XML} document with additional GPU information.}\label{lst:XMLoutput}
\end{listing}

\newpage
\section{Listings of \texttt{diff}s}\label{app:diffs}

\begin{listing}[htbp]
\begin{minted}{diff}
26c26
< our(%scont_config, %scont_part, %scont_jobs, %scont_nodes, %sinfo_cpuinfo);
---
> our(%scont_config, %scont_part, %scont_jobs, %scont_nodes, %sinfo_cpuinfo, @sinfo_gresinfo);
311a312
>     $lrms_cluster->{gres} = [@sinfo_gresinfo];
343a345
>     @sinfo_gresinfo = slurm_read_gresinfo();
533a536,550
> sub slurm_read_gresinfo($){
>     my @sinfo_gresinfo;
>     my $gresinfo;
>     open (SCPIPE,"$path/sinfo -a -h -o \"gresinfo=%G\"|");
>     while(<SCPIPE>){
>         my $string = $_;
>         if ($string !~ m/\(null\)/) {
>             $gresinfo = get_variable("gresinfo", $string);
>             push(@sinfo_gresinfo, $gresinfo);
>         }
>     }
>     close(SCPIPE);
> 
>     return @sinfo_gresinfo;
> }
\end{minted}
\caption{\texttt{SLURMmod.pm}}
\end{listing}

\begin{listing}[htbp]
\begin{minted}{diff}
79c79,80
<         'cpudistribution' => ''
---
>         'cpudistribution' => '',
>         'gres'            => ''
\end{minted}
\caption{\texttt{LRMSInfo.pm}}
\end{listing}

\begin{listing}[htbp]
\begin{minted}{diff}
2983a2984
>             $cmgr->{GeneralResources}{Resource} = $cluster_info->{gres};
\end{minted}
\caption{\texttt{ARC1ClusterInfo.pm}}
\end{listing}

\begin{listing}[htbp]
\begin{minted}{diff}
371a372,374
>         $self->begin('GeneralResources');
>         $self->GeneralResources($data->{GeneralResources});
>         $self->end('GeneralResources');
385a389,393
> }
> 
> sub GeneralResources {
>     my ($self, $data) = @_;
>     $self->properties($data, 'Resource');
\end{minted}
\caption{\texttt{GLUE2xmlPrinter.pm}}
\end{listing}

\begin{listing}[htbp]
\begin{minted}{diff}
204a205
>     std::list<std::string> GeneralResources;
\end{minted}
\caption{\texttt{ExecutionTarget.h}}
\end{listing}

\begin{listing}[htbp]
\begin{minted}{diff}
391a392,397
>     if (!cm.GeneralResources.empty()) {
>                                       out << IString("General resources:") << std::endl;
>       for (std::list<std::string>::const_iterator it = cm.GeneralResources.begin();
>            it != cm.GeneralResources.end(); ++it)
>         out << "  " << *it << std::endl;
>     }
\end{minted}
\caption{\texttt{ExecutionTarget.cpp}}
\end{listing}

\begin{listing}[htbp]
\begin{minted}{diff}
384a385,389
>         if (xComputingManager["GeneralResources"]) {
>           for (XMLNode n = xComputingManager["GeneralResources"]["Resource"]; n; ++n) {
>             ComputingManager->GeneralResources.push_back((std::string)n);
>           }
>         }
\end{minted}
\caption{\texttt{GLUE2.cpp}}
\end{listing}

\end{document}